\documentclass[article, a4paper]{IEEEtran}

\ifCLASSINFOpdf
  \usepackage[pdftex]{graphicx}
\else
\fi

\usepackage{url}
\usepackage{breakurl}
\usepackage[cmex10]{amsmath}
\usepackage{amssymb}
\usepackage{units}
\usepackage{multirow}
\usepackage{rotating}
\usepackage{anyfontsize}
\usepackage{textcomp}

\usepackage[english]{babel}
\usepackage{lipsum}% For 'Lorem Ipsum' dummy text
\usepackage[pscoord]{eso-pic}% The zero point of the coordinate systemis the lower left corner of the page (the default).

\newcommand{\placetextbox}[3]{% \placetextbox{<horizontal pos>}{<vertical pos>}{<stuff>}
  \setbox0=\hbox{#3}% Put <stuff> in a box
  \AddToShipoutPictureFG*{% Add <stuff> to current page foreground
    \put(\LenToUnit{#1\paperwidth},\LenToUnit{#2\paperheight}){\vtop{{\null}\makebox[0pt][c]{#3}}}%
  }%
}%

\begin{document}

\placetextbox{0.5}{1}{This is the author's version of an article that has been published in this journal.}
\placetextbox{0.5}{0.985}{Changes were made to this version by the publisher prior to publication.}
\placetextbox{0.5}{0.97}{The final version of record is available at \href{https://dx.doi.org/10.1109/TII.2016.2641469}{https://dx.doi.org/10.1109/TII.2016.2641469}}%
\placetextbox{0.5}{0.05}{Copyright (c) 2017 IEEE. Personal use is permitted.}
\placetextbox{0.5}{0.035}{For any other purposes, permission must be obtained from the IEEE by emailing pubs-permissions@ieee.org.}%

\title{Experimental Evaluation of Seamless Redundancy Applied to Industrial Wi-Fi Networks%
\thanks{This work was partially supported by the Ministry of Education, University, and Research of Italy (MIUR) in the framework of the Project Wi-Fact ``WIreless FACTory and beyond'' (DM53543). Copyright (c) 2016 IEEE. Personal use of this material is permitted. However, permission to use this material for any other purposes must be obtained from the IEEE by sending a request to pubs-permissions@ieee.org. The authors are with the National Research Council of Italy, Istituto di Elettronica e di Ingegneria dell'Informazione e delle Telecomunicazioni (CNR-IEIIT), I-10129 Torino, Italy (e-mail: gianluca.cena@ieiit.cnr.it, stefano.scanzio@ieiit.cnr.it, adriano.valenzano@ieiit.cnr.it).}}

\author{Gianluca~Cena, \textit{Senior Member}, \textit{IEEE}, Stefano~Scanzio, \textit{Member}, \textit{IEEE}, and\\Adriano~Valenzano, \textit{Senior Member}, \textit{IEEE}}

\maketitle

\begin{abstract}
Seamless redundancy can be profitably exploited to improve predictability of wireless networks in general and, in particular, IEEE 802.11.
According to this approach, packets are transmitted by senders on two (or more) channels at the same time and duplicate copies are discarded by receivers.
As long as the behavior of physical channels is uncorrelated, communication quality improves noticeably, in terms of both transmission latencies and percentage of dropped frames.

In this paper, communication over redundant links has been analyzed by means of a thorough experimental campaign, based on measurements carried out on real devices.
Results confirm that, under typical operating conditions, the assumption of independence among channels in properly designed systems is verified reasonably well.
Indeed, in our experiments, measured link quality indices did not differ more than $10\%$ from what we expected from theory.
This grants for redundant solutions tangible advantages over conventional Wi-Fi networks.
\end{abstract}


\section{Introduction}
All techniques aimed at improving communication reliability on wireless networks through the adoption of medium redundancy rely on the basic assumption that the sources of interference and disturbance on different channels are, to a reasonable extent, independent.
Papers like \cite{2012-WFCS-WoP1}\cite{2012-ETFA-WoP2}\cite{2014-WFCS-WiRed}\cite{2014-ETFA-DDD}\cite{2016-tii-WiRed} showed that, under this hypothesis, reliability of Wi-Fi can be boosted noticeably, to the point that, in some cases, it could be considered as a replacement for cables.
On the other hand, designers of industrial plants are often reluctant to include wireless technologies in their equipment, unless some evidence is provided about their reliability when deployed in factory scenarios.

The large amount of phenomena that may affect the behavior of radio transmissions \cite{2013-TII-Packet_error_rate_rotating_antennas}, and the complexity of the newest IEEE 802.11 versions \cite{2012-std-80211}, make theoretical models inadequate to provide plausible answers.
The same holds, under many respects, for simulation-based studies.
While most network simulators are able to deal satisfactorily with Wi-Fi operation, they usually employ simplified (and generic) models to describe the surrounding environment.
A third approach, complementary to the previous ones, relies on measurements performed on real devices deployed in selected target environments.
Besides Wi-Fi \cite{2007-TII-WLAN}, this method is customarily employed also for other kinds of industrial wireless networks, e.g., those based on timeslotted channel hopping \cite{2015-Sens-Journal} and WirelessHART \cite{2014-tii-Petersen}.
Although the results obtained this way are very specific, nevertheless they can be quite valuable to understand how Wi-Fi behaves ``in the field''.

This paper aims at evaluating the benefits \emph{seamless redundancy} brings when applied to conventional Wi-Fi, and in particular at investigating to what extent the assumption of independence is true in real-world scenarios.
Instead of considering electromagnetic measures as in \cite{2008-TWC-Tanghe}\cite{2015-TWC-Tanghe}, we analyzed the effects of disturbance and interference on applications that communicate over the air.
To this purpose, a set of performance indices has been introduced to characterize the communication quality of a wireless link.
Then, we compared conventional Wi-Fi networks to solutions that exploit seamless redundancy in different operating conditions.

The paper is organized as follows: 
in Section~\ref{sec:Wi-Red} the use of seamless redundancy in wireless networks is briefly described while in Section~\ref{sec:metrics} the metrics we used to compare the performance of simplex and duplex wireless links are introduced.
Section~\ref{sec:testbed} provides details about the testbed on which measurements were carried out and Section~\ref{sec:results} presents and discusses the results obtained in the experimental campaign.
Finally, in Section~\ref{Conclusion} some conclusions are drawn.

\section{Background on Wi-Fi Seamless Redundancy}
\label{sec:Wi-Red}
Seamless redundancy applied to industrial Ethernet is defined in IEC 62439-3 \cite{2012-std-PRP}, also known as  parallel redundancy protocol (PRP).
PRP foresees that two copies of each frame are sent at the same time on a pair of similar (but completely independent) networks.
Each copy is separately delivered to destination, where the copy that arrives first is retained while the second is discarded.
In order to match the frame copies on the two networks, they are unambiguously identified by means of a redundancy control trailer (RCT), added by PRP to each frame.
PRP improves fault tolerance noticeably and, unlike solutions based on network reconfiguration---e.g., the spanning tree protocol (STP)---it also achieves zero-time recovery thanks to fault masking.
This makes it suitable for real-time distributed control systems requiring high availability, like those typically found in  factory automation.

\begin{figure}
  \scriptsize
  \centering
  \includegraphics[width=0.9\columnwidth]{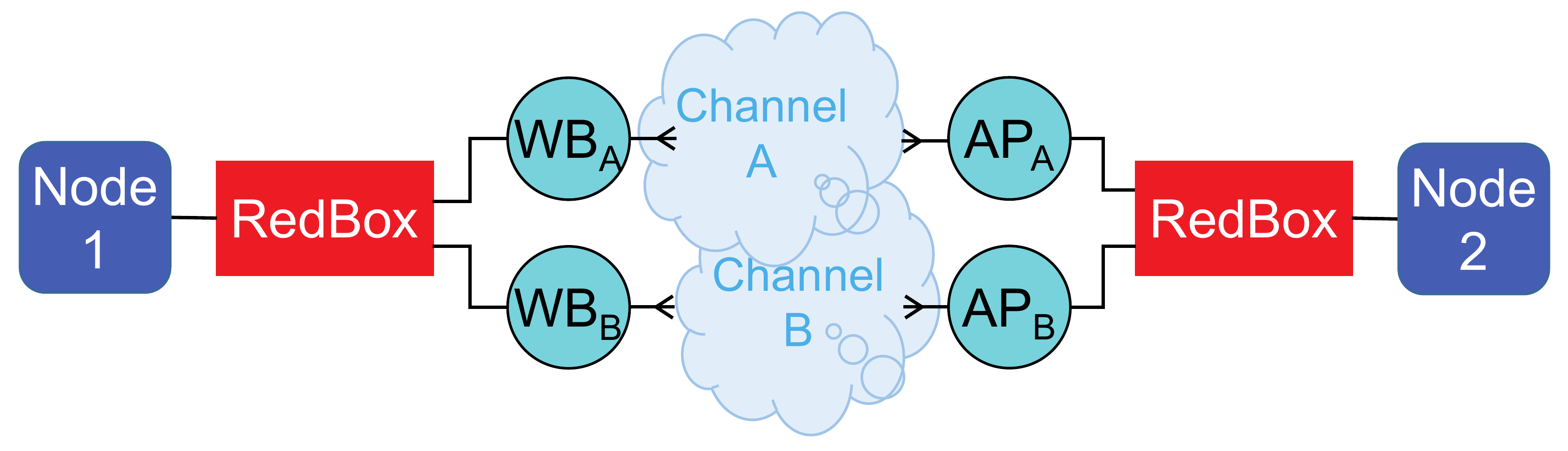}
  \caption{Sample \emph{PRP over Wi-Fi} (PoW) implementation.}
  \label{fig:PoW}
\end{figure}

The idea to apply PRP to Wi-Fi first appeared in \cite{2012-WFCS-WoP1}. 
In that case, identical frame copies were sent on two wireless links operating on different channels.
To this extent, a pair of access points (AP) were used on one side of the redundant link, with which a pair of wireless (client) bridges (WB) were associated on the opposite side, as shown in Fig.~\ref{fig:PoW}.
Special purpose devices, known as RedBoxes \cite{redbox}, took care of managing frame duplication and de-duplication according to PRP.
Besides dealing with permanent failures in wireless network equipment, such an arrangement is also able to face temporary disturbance and interference, which unavoidably affect wireless communication quality.
In the following, the term disturbance will refer to electromagnetic noise and other physical phenomena that cause transmission errors.
Instead, interference denotes the effect of nearby wireless nodes that operate according to the CSMA/CA principle \cite{2014-TII-inteferences} (wireless local area networks, wireless sensor networks, etc.).
Both disturbance and interference lead to a decrease in communication reliability, which impairs the correct system behavior and reduces availability.
A similar approach (iPRP) \cite{2016-TII-IPRP} has been recently applied to wide area IP networks as well.

The solution proposed in \cite{2012-WFCS-WoP1}, we denote \emph{PRP over Wi-Fi} (PoW), has been refined in \cite{2014-WFCS-WiRed}\cite{2014-ETFA-DDD}, but with a slightly different aim.
These papers introduce the \emph{Wi-Fi redundancy} (\mbox{Wi-Red}) protocol, which in its 
basic form closely resembles PoW.
However, Wi-Red is meant to be included in conventional Wi-Fi equipment for improving communication quality, and not fault tolerance.
To avoid the twofold increase of the traffic over the air that plain PRP adoption would imply, Wi-Red purposely introduces \emph{duplicate avoidance} (DA) mechanisms.
Besides saving network bandwidth, they increase communication reliability further, while preserving real-time behavior and fairness.
Basic Wi-Red requires that each \emph{redundant station} (RSTA) is provided with two radio blocks (PHY) and two medium access control entities (MAC) \cite{2016-tii-WiRed}.
It is worth remarking that commercial-off-the-shelf simultaneous dual-band equipment is already provided with such hardware resources.
Additionally, a \emph{link redundancy entity} (LRE) is needed in each RSTA, which takes care of transmitting duplicate copies on the sender side and discarding late copies on the receiver side.
A simplified version of LRE can be implemented in firmware in most existing devices.

Unless network traffic is very low, it is unlikely that the transmissions of the copies of the same frame will start on the two channels at exactly the same time.
This permits solutions based on seamless redundancy to exploit (to a certain degree) \emph{time} diversity---besides \emph{frequency} diversity---which makes them more resilient to short-duration wideband noise pulses.
For any given frame it is possible to identify, on the sender side of the redundant link, a first and a second copy, based on the actual times when their transmissions on air begin.
The same holds for the receiver side, depending on arrival times.
Importantly, order may differ on the two sides.

\section{Metrics about Communication Quality}
\label{sec:metrics}
Two very important performance indices when dealing with wireless communications in industrial environments are the \emph{packet loss ratio} and the \emph{transmission latency}.
The former is the fraction of packets that are definitely lost during transmission and never reach destination.
The latter, instead, applies to successfully delivered packets and represents the time elapsing from the instant when the transmission request is issued on the source node and the instant when the packet is received on the destination.
Both packet losses and transmission latencies worsen communication quality and impair behavior of real-time control systems.

In the following, details are provided about the evaluation of the above performance indices starting from samples measured on an experimental setup which exploits seamless redundancy.
Analysis will focus on a unidirectional point-to-point link between two nodes.
In fact, although traffics produced on the channels of a redundant network by competing RSTAs are not independent, concurrent access to the medium can be regulated by using deterministic MAC overlays,
so that at any time there is only one ongoing transmission in the network.

Every trial consists in a sequence of packets, sent at a regular pace from a source node to a destination node.
The term ``\emph{packet}'' denotes a unit of data produced (or consumed) at the application level (the measurement task, in our case).
Because of retransmissions, more than one IEEE 802.11 MAC frame related to the same packet may be exchanged on any channel.
We will restrict our investigation to wireless systems with duplex redundancy, like PoW.
Physical Wi-Fi channels are referred to as $A$ and $B$ ($C$, in the $\unit[5]{GHz}$ band), while the redundant link is denoted $A\!B$ (or $AC$).
Superscripts are used to distinguish between quantities related to different channels.

\subsection{Packet Loss Ratio}
Let $N$ be the number of packets generated by the source node for a given experiment, while $N_T^X$ and $N_R^X$ are the numbers of packets transmitted and successfully received on a generic channel $X$, respectively.
Moreover, let $M_i$ ($i=1...N$) denote the $i$-th packet generated on the sender side.
Since in basic seamless redundancy schemes, like PoW, every packet is sent on all channels, $N_T^X=N,\forall X\in\{A,B, \mathit{A\!B}\}$.
Concerning delivered packets, $N \geq N_R^{A\!B}\geq \operatorname{max}(N_R^A, N_R^B)$.

The packet loss process for a given packet stream on channel $X$ is defined as a sequence $\left\{l^X_i:i=1...N\right\}$ where $l^X_i$ is an indicator variable
that is true if and only if packet $M_i$ was lost on $X$ \cite{2000-ATM-PLOSS}.
With a slight abuse of notation, $l^X_i=0$ means that the packet was successfully delivered whereas $l^X_i=1$ indicates that it was dropped.
By doing so, the number of dropped packets on channel $X$ is equal to 
\begin{equation}
N_L^X\triangleq N-N_R^X=\sum_{i=1...N} l_i^X
\end{equation}
When seamless redundancy is exploited, a packet is actually lost only when its copies are dropped on every channel. 
This means that $N_L^{A\!B}=\sum_{i=1...N} l_i^A \cdot l_i^B$.

The \emph{packet loss ratio} on channel $X$, referred to a particular experiment, is defined as:
\begin{equation}
\Upsilon_{L}^X\triangleq N_L^X/N = 1 - N_R^X/N
\end{equation}

\subsection{Transmission Latency}\label{sub:transmission_latency}
Let $t_{T,i}^X$ and $t_{R,i}^X$ be the times when packet $M_i$ is sent by the source node and received by the destination node on channel $X$, respectively.
It is worth remembering that packets for which $l_i=1$ never arrive, and so $t_{R,i}^X$ is not defined for them.
The \emph{transmission latency} incurred by a packet $M_i$ correctly delivered on channel $X$ is defined as:
\begin{equation}
d^X_i\triangleq t_{R,i}^X-t_{T,i}^X
\end{equation}
To ease reasoning, packets that were dropped may be thought of as having infinitely large latency ($l_i^X=1$ implies $d^X_i=\infty$).

Since both copies of packet $M_i$ are sent at about the same time, we can safely assume that $t_{T,i}^{A} \simeq t_{T,i}^{B} \simeq t_{T,i}^{A\!B}$.
For instance, in our testbed they never differed by more than $\unit[90]{\mu s}$ ($\unit[1.25]{\mu s}$, in absolute terms, on average).
Moreover, $t_{R,i}^{A\!B}=\operatorname{min}(t_{R,i}^{A},t_{R,i}^{B})$ because only the copy which arrives first is retained in PoW.
This means that the latency on the redundant link can be satisfactorily evaluated as $d_i^{A\!B}=\operatorname{min}(d^A_i,d^B_i)$.

\section{Experimental Evaluation}
\label{sec:testbed}
In \cite{2014-WFCS-WiRed}\cite{2014-ETFA-DDD}\cite{2016-tii-WiRed}, disturbance and interference on the two channels of a redundant Wi-Fi link were modeled as independent stochastic processes.
In order to complement the simulation campaigns carried out in those papers, we performed a thorough experimental evaluation aimed at assessing to which extent that assumption was acceptable.
Only the basic Wi-Red case---i.e., without duplicate avoidance---has been considered.
Its behavior closely resembles PoW, the only difference being that redundancy in Wi-Red is dealt with at the data-link layer (and not above).
DA mechanisms were purposely neglected, because this paper focuses on the properties of wireless channels and not on a specific redundancy approach.

Instead of relying on the full PoW implementation depicted in Fig.~\ref{fig:PoW}, we adopted the simpler approach shown in Fig.~\ref{fig:testbed}.
Concerning the evaluation of the degree of independence between channels, both of them provide exactly the same results.
Basically, sequences of packets were generated according to a periodic pattern and jointly exchanged on two distinct channels, in the same way as for PoW.
At the same time, information was gathered about transmission and reception times. 
Very large log files were collected during the experiments that, on the whole, lasted several months.

A subsequent post-processing phase permitted to characterize several aspects concerning the wireless spectrum, and in particular the independence of separate Wi-Fi channels.
This approach permits to prevent some overheads that would otherwise affect the measured quantities, e.g., the delays introduced by the online software de-duplication algorithm.

Besides latencies, dropped packets were also considered.
We assume that any packet which is not received by the end of the experiment is actually lost.
Since reception remained enabled for $\unit[5]{s}$ after the end of transmissions, this approximation is acceptable.
For space reasons, only the most significant results are included and discussed in the following.

\subsection{Testbed}

\begin{figure}
  \scriptsize
  \centering
  \includegraphics[width=0.9\columnwidth]{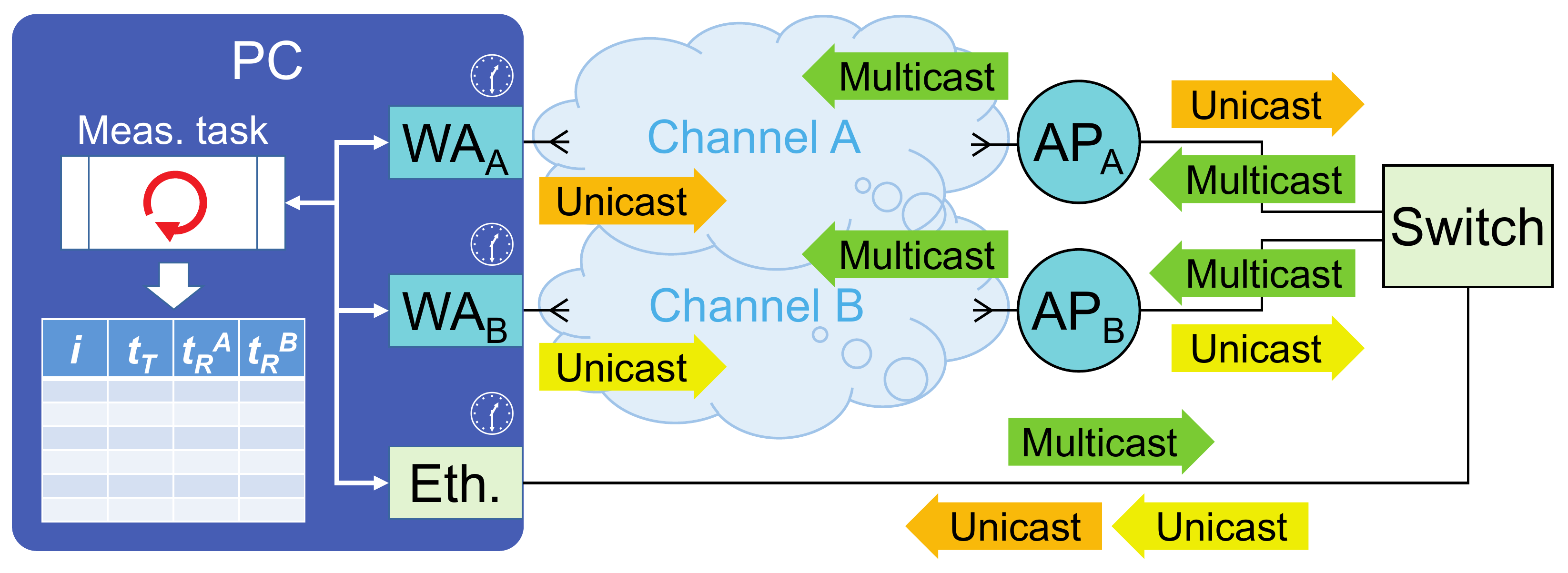}
  \caption{Testbed used for experimental measurements.}
  \label{fig:testbed}
\end{figure}

Our experimental testbed consists of a PC provided with two wireless adapters (WA) and two APs (see Fig.~\ref{fig:testbed}).
The latter are connected to the PC via Ethernet by means of a Netgear ProSafe GS105E switch with a link transfer rate of $\unit[1]{Gbps}$. 
The Ethernet link between the switch and the PC runs at full speed, whereas the links between the switch and the APs may be limited by the speed of the Ethernet interfaces of the latter (e.g., $\unit[100]{Mbps}$).
The PC is equipped with an Intel i3-4150 CPU running at $\unit[3.5]{GHz}$ and $\unit[4]{GB}$ of RAM, and runs Linux (kernel v.~3.13).
Wi-Fi adapters are managed by the \texttt{ath9k} driver.
Each of them was associated to a distinct AP, in order to set up two independent communication links.

Since the PC takes the roles of both transmitter and receiver, its time base can be exploited to evaluate time differences.
Interference among different pieces of hardware (wired/wireless adapters) and software (operating system and local tasks) inside the PC is mostly negligible.
In fact, in-node delays are greater than $\unit[100]{\mu s}$ and, $99.99\%$ of the times, less than $\unit[300]{\mu s}$, whereas the $99.99$ percentile on measured latencies can be as high as $5$ to $\unit[100]{ms}$, depending on the experiment.

A task running on the PC generates a periodic stream of packets, and sends them on both channels at the same time by using raw POSIX sockets.
The same \emph{sequence number} (corresponding to the index $i$ used in the following) was included in both copies of the same packet, in order to identify uniquely each piece of information exchanged over the air (similarly to RCT in PRP).
Packets are received on the other side of the links.
Timestamps were taken on each transmission and reception, stored in a data structure held in main memory (RAM), and dumped to a file only at the end of each experiment.
Duplicate copies were not removed at runtime.
Instead, we dealt with them in the post-processing phase.
This is appropriate because, unlike Wi-Red, PoW does not exploit in any way the outcome of exchanges to drive frame transmissions (there is no duplicate avoidance).

In experiments involving only the industrial, scientific, and medical (ISM) band at $\unit[2.4]{GHz}$, two D-Link DWA-556 were chosen as Wi-Fi adapters and two TP-Link TL-WA801ND as APs.
Although all of them comply to IEEE 802.11n, they were set to operate according to IEEE 802.11g, in order to keep subsequent frame transmissions as separate as possible (besides, possibly, queuing phenomena).
This is clearly impossible when frame aggregation and block acknowledgment are enabled.
Link speed varied over time, because rate adaptation was enabled in all the experiments.
The maximum achieved speed was $\unit[54]{Mbps}$.
We configured the two APs to operate on different non-overlapping channels (channels $1$ and $11$ in the $\unit[2.4]{GHz}$ band), denoted $A$ and $B$ for short.
Since channels are adequately spaced, hardly they are jointly affected by transmissions of other nodes---including those which exploit channel bonding ($\unit[40]{MHz}$) or operate on non-canonical channels (other than $1$, $6$, or $11$).

For experiments involving also the $\unit[5]{GHz}$ band, dual-band TP-Link TL-WDN4800 wireless adapters were employed.
Although only one link was actually set to operate in the $\unit[5]{GHz}$ band (on channel 44), termed as $C$ in the following, identical wireless adapters were used in each experiment, in order to minimize discrepancies different interfaces to the PC system bus may cause.
The AP used for channel $C$ was a dual-band Linksys WRT320N, set to comply with IEEE 802.11a.

\subsection{Environment characterization}
Quality of wireless communication heavily depends on the surrounding environment.
Since the lab where experimentation was performed is located very close to another lab where research on wireless networks was also carried out, the radio environment was not under our control.
What is worse, conditions were continuously changing, which implies that repeatability of experiments could not be ensured.

\begin{figure}
  \scriptsize
  \centering
  \includegraphics[width=0.8\columnwidth]{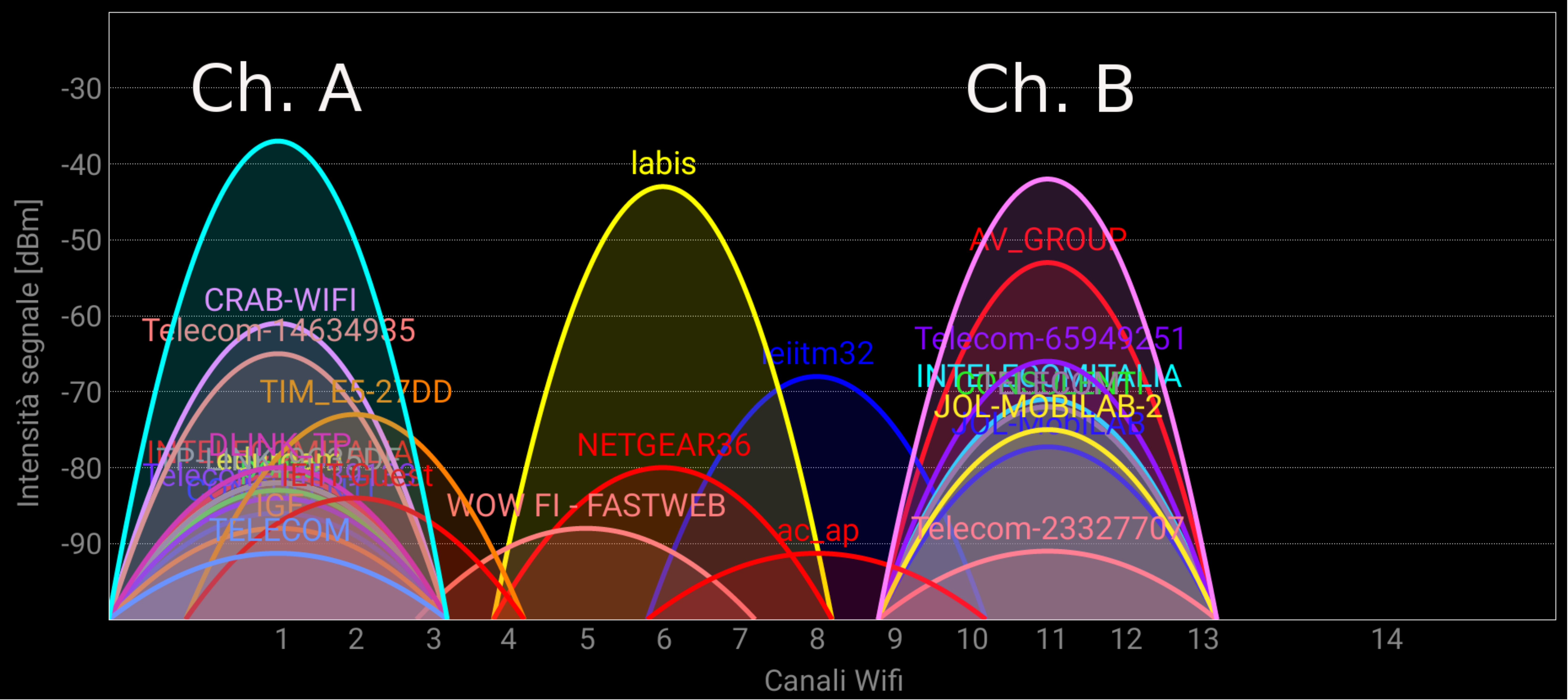}
    \begin{center}
    \tabcolsep=0.05cm
    \scriptsize
    \begin{tabular}{l||c|c|c}
      Parameter                      & Ch. $A$                     & Ch. $B$                    & Ch. $C$   \\
      \hline \hline
      Band                             & $\unit[2.4]{GHz}$         & $\unit[2.4]{GHz}$          & $\unit[5]{GHz}$ \\
      Channel                          & $1$ ($\unit[2.412]{GHz}$) & $11$ ($\unit[2.462]{GHz}$) & $44$ ($\unit[5.220]{GHz}$) \\
      \hline
      Number of BSSs                   & $16 / 16 / 14$            & $22 / 16 / 15$             & $2 / 2 / 3$             \\
      Number of STAs                   & $502 / 333 / 1879$        & $463 / 250 / 2097$         & $81 / 20 / 252$         \\
      Tx. rate [$\unit[]{pkt/s}$]      & $202 / 141 / 157$         & $494 / 61 / 333$           & $22 / 20 / 31$          \\
      Tx. rate [$\unit[]{Mbps}$]       & $0.335 / 0.292 / 0.331$   & $1.452 / 0.222 / 0.473$    & $0.036 / 0.032 / 0.067$ \\
      Multicast [$\unit[]{\%}$]        & $39.5 / 65.7 / 59.4$      & $15.6 / 45.3 / 19.7$       & $94.4 / 99.5 / 95.8$    \\
      \hline
    \end{tabular}
  \end{center}
  \caption{Nearby BSSs in the $\unit[2.4]{GHz}$ band (top) and characteristics of the interfering traffic on the channels used in the testbed (bottom). 
    Notation $\cdot/\cdot/\cdot$ refers to the values obtained in the $3$ different WireShark acquisitions.}
  \label{fig:networks}
\end{figure}

We inspected the traffic in the radio environment when the measurement task was switched off.
The screenshot in the upper part of Fig.~\ref{fig:networks} shows, as an example, the nearby BSSs that were visible at some point in time in the ISM band.
As can be seen, some tens APs were concurrently active.
Information in the lower part of the figure was obtained by running the WireShark network protocol analyzer for one hour on every channel we used in the experiments ($A$, $B$, and $C$).
Each sampling was repeated three times, in different days and by varying the time of day.
Results show that variability of channel behavior over time is significant, especially on channel $B$.
The only thing we can infer from them is that, generally speaking, traffic on the $\unit[5]{GHz}$ band was noticeably lower than on the $\unit[2.4]{GHz}$ band.
Erraticness of the wireless spectrum is further confirmed by the measurement campaign.
In fact, as will be shown in Section~\ref{sec:results}, repeating the same experiment in different days typically led to quite different results.

For these reasons, each trial described in the following has to be considered as standing on its own, and should not be compared to other trials.
In other words, results are not meant to characterize the ``typical'' behavior of physical wireless channels.
On the contrary, since experiments involved the redundant link and both of its channels at the same time, measurements can be used to determine the effectiveness of seamless redundancy approaches over conventional \mbox{Wi-Fi}.
In particular, each single trial permits to compare, in a reliable and meaningful way (i.e., in specific yet identical environmental conditions), the behavior of either of the physical channels to a link that applies PRP on top of them.

\subsection{Experimental configurations}
Two configurations were adopted, to analyze \emph{unicast} and \emph{multicast} packets, respectively.
Unicast packets are quite reliable but may suffer from larger transmission jitters because of retransmissions and exponential backoff.
Multicast packets are required to properly enable producer-consumer schemes, but are noticeably slower since packets are typically transmitted by the APs using the lowest available bit rate.

As depicted in Fig.~\ref{fig:testbed}, unicast packets are sent directly by the PC using its wireless adapters.
They are received by the related AP, which relays them to the switch and then back to the Ethernet port of the PC.
Serialization carried out by the switch on its egress port may introduce small errors (less than $\unit[1]{\mu s}$) on packet arrival times, which are mostly irrelevant for our measurements.
They typically affect the slowest copy, which is discarded by seamless redundancy schemes.

Directions were inverted for multicast transmissions. 
In this case, packets are sent via Ethernet from the PC to the switch, from which they reach the two APs and are then relayed back to the PC through its wireless adapters.
This change is necessary since, in the hop from a station (STA) to the AP, multicast frames use acknowledged transmissions.

\subsection{Improving channel behavior}\label{sub:improving_channel_behavior}
Preliminary tests highlighted two specific issues, which must be dealt with properly since they
cause packet delays and losses on both channels according to predictable schemes.

The first is due to a Linux service known as the \emph{network manager}.
One of its tasks is to periodically scan Wi-Fi channels in order to provide the STA with a list of APs it can associate to.
Besides dealing with device mobility (by enabling roaming), the network manager also permits to tackle cases where the association between the STA and the AP is lost, due to errors that cause the relevant timeouts to expire.
When channel scan is being carried out by an adapter, it cannot send or receive the application packets, hence temporarily impairing communication.
A straightforward solution is to disable the network manager \cite{2014-ETFA-trama}\cite{2016-ETFA-WiFi}, but doing so makes the STA unable to reassociate to the AP.
A better approach is to develop a new network management service, specifically tailored to the needs of industrial applications. 
For instance, if node mobility is not required and the Wi-Fi network topology is fixed, the simplest solution is a process that automatically reconnects a STA to the AP as soon as disconnection is detected (this can be achieved, e.g., by analyzing the \texttt{proc} filesystem).

The second issue has to do with the transmission of multicast packets, as managed by the IEEE 802.11 mechanism based on delivery traffic indication messages (DTIM).
If a device that can enter a power-saving state (e.g., mobiles) associates to the AP, the AP is forced to buffer all outgoing multicast frames and send them following beacons, according to DTIM rules.
This causes undue delays for real-time multicast packets.
A simple remedy is to make the SSID hidden, so that STAs other than those involved in the system are prevented from accidentally associate to the AP.
A more effective approach \cite{2016-ETFA-WiFi} is to use authentication, as per the \mbox{Wi-Fi} protected access (WPA).

\section{Experimental Results}
\label{sec:results}
In order to characterize some statistical properties of redundant Wi-Fi links, we carried out some experimental campaigns using the above testbed.
Two kinds of environments were taken into account, namely a real industrial plant and our laboratory.

\subsection{Industrial environment}

\begin{table}
  \caption{Comparison between measurements in a real industrial plant and in our laboratory (network manager enabled).}
  \label{tab:industrial_responsiveness}
  \scriptsize
  \fontsize{6.0}{7}
  \begin{center}
  \tabcolsep=0.13cm	
   \begin{tabular}{cc|cccc|cc|c}
      
                        Env. & Chan.
			& $\overline{d}$$^X$ & $\sigma_{d}^X$ & $d^X_{p99.99}$  & $d^X_{\operatorname{Max}}$
			& $\Upsilon_{d>3\mathrm{ms}}^X$ 
			& $\Upsilon_{d>10\mathrm{ms}}^X$ 
			& $\Upsilon_{L}^X$ \\
			& ($X$)
			& \multicolumn{4}{c|}{($\unit[]{ms}$)} 
                        & \multicolumn{2}{c|}{(\textperthousand)}
			& \multicolumn{1}{c}{(\textperthousand)} \\
      \hline \hline
      \multirow{3}{*}{\begin{sideways}Plant\end{sideways}} 
      & $A$  & 1.42 & 10.41 & 447.2 & 2928 & 44.67 & 10.56 & 0.026 \\
      & $B$  & 1.14 &  3.26 &  71.68 &  367.7 & 26.62 & 8.48  & 0.000 \\
      & $A\!B$ & 0.87 &  2.26 &  54.81 &   61.96 &  7.92 & 4.90  & 0.000 \\
      \hline
      \multirow{3}{*}{\begin{sideways}Lab\end{sideways}} 
      & $A$  & 1.97 & 6.31 & 117.3 & 174.1 & 93.60  & 22.99 & 0.000 \\
      & $B$  & 2.53 & 6.91 & 119.4 & 781.5 & 157.4 & 39.44 & 0.0093 \\
      & $A\!B$ & 1.38 & 5.20 & 111.9 & 119.9 & 31.15 & 13.80 & 0.000 \\
      \hline \hline
    \end{tabular}
  \end{center}
\end{table}

Experiments in the industrial plant, which included many robotic cells, were potentially affected by heavy disturbance due, for the most part, to sparks produced by high-current welding guns.
We left the network manager enabled, as these experiments were unattended and we had to be sure that disconnections of the STAs from the APs were not permanent.
Because of the implementation of this service, its interference occurred simultaneously on the two channels, which is the worst condition for seamless redundancy.

Each experiment lasted $\unit[6]{days}$ and the measurement period $T_c$ was set to $\unit[100]{ms}$.
As outcome, a sequence of tuples $\langle l^X_i, d^X_i \rangle$, i.e., loss and latency of each single packet $M_i$, was produced for every channel $X$.
Typical statistical indices were calculated from these samples.

\begin{table*}
  \caption{Experimental results concerning transmission latency and packet loss ratio\hspace{\textwidth}($\unit[2.4]{GHz}$ band, network manager disabled).}
  \label{tab:latency}
  \scriptsize
  \fontsize{6.0}{7}
  \begin{center}
  \tabcolsep=0.12cm	
    
    \begin{tabular}{cccc|cccc|cccccccc|c|cc}
      
      \multicolumn{2}{c}{Type} & $T_c$ & Chan. 
			& $\overline{d}$$^X$ & $\sigma_{d}^X$ & $d^X_{p99.99}$  & $d^X_{\operatorname{Max}}$ 
			& $\Upsilon_{d>1\mathrm{ms}}^X$ & $\hat{\Upsilon}_{d>1\mathrm{ms}}^{A\!B}$
			& $\Upsilon_{d>3\mathrm{ms}}^X$ & $\hat{\Upsilon}_{d>3\mathrm{ms}}^{A\!B}$
			& $\Upsilon_{d>10\mathrm{ms}}^X$ & $\hat{\Upsilon}_{d>10\mathrm{ms}}^{A\!B}$
			& $\Upsilon_{d>30\mathrm{ms}}^X$  & $\hat{\Upsilon}_{d>30\mathrm{ms}}^{A\!B}$
                        & $D_{KS}$
			& $\Upsilon_{L}^X$ & $\hat{\Upsilon}_{L}^{A\!B}$ \\
			&  & ($\unit[]{ms}$) & ($X$)
			& \multicolumn{4}{c|}{($\unit[]{ms}$)} 
			& \multicolumn{8}{c|}{(\textperthousand)} 
                        &
			& \multicolumn{2}{c}{(\textperthousand)} \\

      \hline \hline

      \multirow{12}{*}{\begin{sideways}Unicast\end{sideways}}  & \multirow{6}{*}{\begin{sideways}Weekend\end{sideways}} & 
              & $A$  & 1.12  & 1.04  & 19.63  & 41.43    & 231.4   & -- 	  & 34.04 & -- 	  & 4.373 & -- 	  & 0.005  & --     &   & 0.000  & --   \\
       & & 10 & $B$  & 1.01  & 1.14  & 32.22  & 79.06    & 174.4   & -- 	  & 37.54 & -- 	  & 2.320 & -- 	  & 0.126  & --     &   & 0.000  & --   \\
       & &    & $A\!B$ & 0.71  & 0.14  &  4.06  & 14.03    & 22.93   & 40.37& 0.429 & 1.278 & 0.003 & 0.010 &0.000  & 0.000           & 0.102 & 0.000  & 0.000   \\
      \cline{3-19}
       & &     & $A$  & 0.92  & 0.92  & 26.80  & 100.9  & 270.1 & --     & 13.39  & -- 	   & 0.227 & -- 	   & 0.090  & --    &   & 0.000  & --   \\
       & & 100 & $B$  & 0.94  & 0.86  & 12.86  & 101.0  & 182.6 & -- 	& 5.521  & -- 	   & 0.250 & -- 	   & 0.054  & --    &   & 0.000  & --   \\
       & &     & $A\!B$ & 0.72  & 0.10  &  2.77  &  6.25  & 17.38 & 49.32 & 0.058  & 0.074   & 0.000 & 0.000 & 0.000  & 0.000         & 0.143 & 0.000  & 0.000 \\
      \cline{2-19}
        & \multirow{6}{*}{\begin{sideways}Weekdays\end{sideways}}  & 
              & $A$  & 0.90  & 0.59  & 16.89  & 56.72    & 272.6 & -- 	  & 10.94 & -- 	   & 0.558 & -- 	   & 0.020  & --    &   & 0.000  & --   \\
       & & 10 & $B$  & 0.91  & 0.71  & 15.53  & 123.3   & 151.2 & -- 	  & 18.50 & -- 	   & 0.664 & -- 	   & 0.006  & --    &   & 0.000  & --   \\
       & &    & $A\!B$ & 0.69  & 0.10  &  3.22  & 12.35    & 11.38 & 41.22 & 0.149  & 0.202 & 0.001 & 0.000 & 0.000  & 0.000          & 0.078 & 0.000  & 0.000   \\
      \cline{3-19}
       & &     & $A$  & 0.92  & 0.96  & 28.20  & 100.8  & 281.9 & -- 	  & 11.01 & -- 	   & 0.529 & -- 	   & 0.093  & --    &   & 0.000  & --   \\
       & & 100 & $B$  & 0.95  & 1.02  & 19.17  & 100.9  & 191.3 & -- 	  & 19.17 & -- 	   & 0.733 & -- 	   & 0.059  & --    &   & 0.000  & --   \\
       & &     & $A\!B$ & 0.71  & 0.11  &  3.47  &  8.93  & 16.32  & 53.92 & 0.204 & 0.211  & 0.000 & 0.000 & 0.000  & 0.000          & 0.084 &  0.000 & 0.000   \\

      \hline \hline

      \multirow{12}{*}{\begin{sideways}Multicast\end{sideways}} & \multirow{6}{*}{\begin{sideways}Weekend\end{sideways}} & 
              & $A$  & 1.31  & 1.19  & 25.85  & 69.54   & 225.1  & -- 	& 63.03 & -- 	   & 3.431 & --    & 0.741   & --  &   & 0.698  & --   \\
       & & 10 & $B$  & 1.93  & 2.12  & 33.31  & 60.77   & 479.0  & -- 	& 158.6 & -- 	   & 14.99 & --    & 2.101   & --  &   & 1.910  & --   \\
       & &    & $A\!B$ & 1.02  & 0.40  & 8.46   & 36.29   & 104.8  & 107.8& 9.558  & 9.999 & 0.053 & 0.051 & 0.003 & 0.002   & 0.030 & 0.0028 & 0.0013  \\
      \cline{3-19}
       & &     & $A$  & 1.41  & 1.24  & 24.27  & 55.37  & 260.2  & -- 	  & 81.18 & -- 	  & 3.203 & -- 	   & 0.765    & --  &   & 0.727  & --   \\
       & & 100 & $B$  & 1.78  & 1.81  & 28.40  & 55.05  & 432.6  & -- 	  & 137.1 & -- 	  & 14.42 & -- 	   & 5.788    & --  &   & 5.727  & --   \\
       & &     & $A\!B$ & 1.04  & 0.40  & 7.99   & 19.43  & 110.8  & 112.6  & 10.06 & 11.13 & 0.062 & 0.046 & 0.029    & 0.004& 0.031 & 0.029 & 0.0042 \\
      \cline{2-19}
        & \multirow{6}{*}{\begin{sideways}Weekdays\end{sideways}} & 

              & $A$  & 1.65  & 2.70  & 90.01  & 383.4 & 330.4 & -- 	& 107.1 & --      & 11.35 & --      & 2.903  & --   &   & 1.919 & --   \\
       & & 10 & $B$  & 2.51  & 3.26  & 51.72  & 137.7 & 566.6 & --       & 232.5 & --      & 43.91 & --      & 11.20 & --   &   & 9.948 & --  \\
       & &    & $A\!B$ & 1.15  & 0.80  & 18.08  & 105.3 & 196.6 & 187.2   & 29.64 & 24.89 &  0.852 & 0.498  & 0.044  & 0.032  & 0.022 & 0.027  & 0.019  \\
      \cline{3-19}
       & &     & $A$  & 1.57  & 1.63  & 31.25  & 106.3  & 321.1  & -- 	 & 109.7& -- 	& 11.09& --     & 5.203 & --    &   & 5.091   & -- \\
       & & 100 & $B$  & 2.16  & 2.58  & 39.25  & 92.71  & 501.8  & -- 	 & 192.9& -- 	& 32.09& --     & 11.52 & --    &   & 11.05   & -- \\
       & &     & $A\!B$ & 1.12  & 0.63  & 13.86  & 44.35  & 169.4  & 161.1 & 22.98 & 21.15& 0.506& 0.356 & 0.174 & 0.060  & 0.035 & 0.171   & 0.056 \\
      \hline \hline

    \end{tabular}
  \end{center}
\end{table*}

Results for each experiment are described by three contiguous rows in the tables, related to channels $A$, $B$, and $A\!B$, respectively.
Operating conditions are reported in the leftmost part of the table.
The second set of columns concerns the transmission latency of packets that were actually received (i.e., those with $l_i^X=0$).
In particular, the mean value ($\overline{d}$$^X$), standard deviation ($\sigma_{d^X}$), 99.99 percentile ($d^X_{p99.99}$), and maximum (worst-case) value ($d^X_{\operatorname{Max}}$) are included.

The third set of columns is about the \emph{deadline miss ratio} ($\Upsilon_{d>H}^X$), which corresponds to the fraction of transmitted packets whose latency exceeded a given threshold $H$ (including the dropped ones), that is:
\begin{equation}
\Upsilon_{d>H}^X \triangleq 
  \frac{N_L^X + N_{R\mid d>H}^X}{N} =
  \Upsilon_{L}^X+(1-\Upsilon_{L}^X)\cdot \frac{N_{R\mid d>H}^X}{N_{R}^X}
\label{eq:dmr}
\end{equation}
where $N_{R\mid d>H}^X$ is the number of received packets with $d>H$.
For instance, $\Upsilon_{d>3\mathrm{ms}}^X$ refers to packets which missed the $\unit[3]{ms}$ deadline.
Finally, in the rightmost part of the table, the packet loss ratio $\Upsilon_{L}^X$ (which can be seen as $\Upsilon_{d=\infty}^X$) is shown.

Latencies for unicast packets in a representative experiment have been reported in the upper part of Table \ref{tab:industrial_responsiveness}. 
Channel $A$ performed quite badly: measured worst-case latency exceeded $\unit[2.9]{s}$ and some packets went lost in spite of retransmissions ($\Upsilon_{L}^A=\unit[0.026]{}$\textperthousand).
This was likely due to temporary disconnections of the related adapter from the AP (overcome by the network manager).
Channel $B$ performed better, however its worst-case latency was as high as $\unit[0.368]{s}$.
On the contrary, the latency for the redundant link $A\!B$ never exceeded $\unit[62]{ms}$, shorter than $T_c$ and small enough for several factory automation contexts.

\subsection{Lab environment}
\label{sec:lab}
We repeated the same experiment as the industrial plant in our lab as well (in particular, the network manager was left on).
Results are reported in the lower part of Table \ref{tab:industrial_responsiveness}.
As can be seen, the two environments were not substantially dissimilar: in fact, welding is known to generate electromagnetic noise far below $\unit[2.4]{GHz}$ \cite{2006-ABB-WISA-disturb}\cite{1997-IECON-disturb}, which means that its effect on wireless communication is negligible.
Clearly, several real industrial scenarios exist where the level and kind of disturbance may be so high to impair Wi-Fi communications severely. 
However, it is not the aim of this paper to find and characterize such environments.
Besides, using ``worst cases'' as typical operating conditions would probably be misleading.

Overall, the industrial plant we took into account did not behave noticeably worse than our lab.
For this reason, the following analysis focuses on the lab case only.
So as to avoid the issues highlighted in Section \ref{sub:improving_channel_behavior}, we permanently disabled the network manager.
In the same way, the SSIDs of the APs were hidden, so that extraneous devices can not associate.
In this way, the only causes of interference and disturbance depended on the surrounding environment.

We carried out a first set of lab experiments by varying the packet transmission type (unicast/multicast), kind of day (weekdays/weekend), and measurement period ($\unit[10]{ms}$ or $\unit[100]{ms}$).
In the case $T_c=\unit[10]{ms}$, much more samples could be acquired, which improves statistical relevance of data.
Checking results against those obtained with $T_c=\unit[100]{ms}$ permits to assess whether $\unit[10]{ms}$ is long enough to prevent the occurrence of queuing phenomena inside devices or channel congestion, which would make results less appealing as we are interested in typical disturbance and interference conditions over the air.
Each trial lasted exactly one day.
This means that $864000$ tuples per trial were collected for each channel when $T_c=\unit[100]{ms}$, and ten times more when $T_c=\unit[10]{ms}$.

Results are reported in Table~\ref{tab:latency} which, for the redundant link, additionally shows the estimates $\hat{\Upsilon}_{d>H}^{A\!B}$ of the deadline miss ratio and $\hat{\Upsilon}_{L}^{A\!B}$ of the packet loss ratio (see Sections \ref{sec:DMR} and \ref{sec:PLR}).
As can be noted, seamless redundancy always boosts performance over single channels.
Concerning latencies, $\overline{d}$$^X$ only shrinks slightly.
Conversely, improvements on $\sigma_{d^X}$, $d^X_{p99.99}$, and $d^X_{\operatorname{Max}}$ are more tangible (up to a tenfold decrease).
This means that network predictability gets better.
Delays of unicast packets in PoW never practically exceeded $\unit[10]{ms}$, as shown in the column $\Upsilon_{d>10\mathrm{ms}}^{A\!B}$ of the table.
However, this does not hold for physical channels, which means that they occasionally incurred in packet queuing when $T_c=\unit[10]{ms}$.
Delays of multicast packets were slightly larger than unicast ones.
This is probably due to the lower bit rate used by the former, which was predominant over the (not too frequent) frame retransmissions carried out by the latter.

No packets were dropped for unicast (acknowledged) transmissions, because of the quite effective retransmission mechanism foreseen by the IEEE 802.11 MAC.
Conversely, when multicast (unacknowledged) transmissions are considered, up to \unit[$11.05$]{\textperthousand} of them were definitely lost on physical channels, whereas losses never exceeded \unit[$0.171$]{\textperthousand} for PoW.
Hence, the improvements seamless redundancy achieved for $\Upsilon_L$ were significant (more than one order of magnitude).

\subsection{Correlation over time}
We analyzed channels $A$ and $B$ separately, to determine how much distinct packet transmissions were correlated.

\subsubsection{Packet Loss Ratio}
Let $P_{l_i^X}\triangleq P(l_i^X=1)$ be the probability that packet $M_i$ is dropped on channel $X$. The expected value of the packet loss ratio on $X$ for a given transmission sequence is equal to $\operatorname{E}[\Upsilon_L^X]=\frac{1}{N}\sum_{i=1...N}P_{l_i^X}$.
If frame transmissions on the same channel are independent from each other and the probability $P_{l_i^X}$ does not depend on $i$, they can be modeled as Bernoulli trials.
In this case, let $L^X$ be a random variable that models the outcome (success/failure) of any single transmission. 
Then, $\operatorname{E}[\Upsilon_L^X]=P_{L^X}$.
Under the above hypothesis, since experiment durations were quite long, the approximation $P_{L^X}\simeq \Upsilon_L^X$ holds reasonably well.

\subsubsection{Autocorrelation}
Wireless spectrum characteristics vary indeed over time, due to a plethora of causes that are typically unknown to the system designer.
Transmission of packet $M_i$ is influenced by the overall state of the wireless spectrum at the specific time when the packet is sent.
We are interested in evaluating how much (slow) variations of the spectrum state may affect distinct transmissions, hence making them ultimately correlated.
In particular, we wish to determine to which extent the transmission outcomes for packets $M_{i}$ and $M_{i+k}$ are correlated when $k$ (which is related to the time between the transmissions of the considered packets) is varied.

\begin{figure}[t]
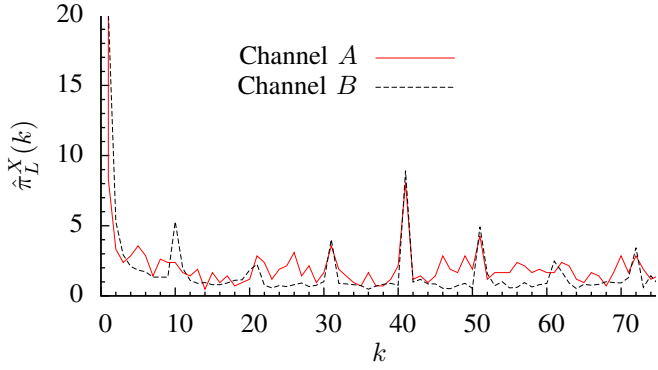

  \centering
  \include{FIG4-16-0457}
  \caption{Plot of $\hat{\pi}^X_{L}(k)$ versus lag $k$ for the two channels
	of redundant link $A\!B$,
	with a \emph{multicast} transmission type, during \emph{weekends}, and with $T_c=\unit[10]{ms}$.}
	\label{fig:independence}
\end{figure}

The \emph{discrete autocorrelation} at lag $k$ for the (binary) packet loss process $\left\{l_i\right\}$ is defined as:
\begin{equation}
	R_{ll}(k) \triangleq \operatorname{E} \left[ l_i \cdot l_{i+k} \right] = P(l_{i}=1 \wedge l_{i+k}=1)
\end{equation}
In the case of our experimental samples $\left\{l^X_i:i=1...N\right\}$, for $k\in [0,K]$ it can be evaluated as:
\begin{equation}
	\hat{R}^X_{ll}(k) = \frac{1}{N-K} \sum_{i=1}^{N-K} \left( l^X_i \cdot l^X_{i+k} \right)
\end{equation}
If $l^X_i$ and $l^X_{i+k}$ are independent, then $R^X_{ll}(k) = P_{l_i^X} \cdot P_{l_{i+k}^X}$ and, in stationary conditions, $R^X_{ll}(k) = {P_{L^X}}^2$.
Thus, the quantity $\hat{\pi}^X_{L}(k) \triangleq {\hat{R}^X_{ll}(k)}/{{\Upsilon_L^X}^2}$ can be used as an indication of the degree of dependence over time (value $1$ means independence).

In Fig. \ref{fig:independence}, $\hat{\pi}^X_{L}(k)$ is plotted versus $k$ for multicast transmissions, during weekend, and with $T_c=\unit[10]{ms}$.
As can be seen, for adjacent samples ($k=1$) it is in the order of $10$ while, when $k$ increases, it converges, on average, to $1$.
A number of peaks can still be found, located about every $10$ samples ($100$, $200$, $\unit[300]{ms}$, etc.).
They are due to beacons of nearby Wi-Fi networks (sent every $\unit[102.4]{ms}$).
In fact, if a beacon frame causes the corruption of a packet, reasonably one of the following beacons might cause a similar interference.

\subsubsection{Error burstiness}
In many industrial applications, losing a single packet is not a severe error, as long as many of them are not dropped in a row.
Let an error burst be defined as a packet sequence starting at $M_s$ and ending at $M_e$ where $(l_{s-1}=0 \vee s=1) \wedge (l_{e+1}=0 \vee e=N) \wedge (l_i=1, \forall i \in \{s...e\})$.
The length of that burst is then given by $B=e-s+1$.

Part of the frequency distribution of experimental burst lengths is shown in the rightmost columns of Table~\ref{tab:responsiveness}.
As can be seen, the number $N_B$ of bursts with length $B$ decreases as $B$ grows up (although not as much as for a Bernoulli process).
Notably, in the experiments where redundancy was exploited, $B$ never exceeded $2$ (see $B_\mathrm{Max}$ in the last column of the table).

\begin{table}
  \caption{Experimental results about error burstiness\hspace{\textwidth}($\unit[2.4]{GHz}$ band, network manager disabled).}
  \label{tab:responsiveness}
  \scriptsize
  \fontsize{6.0}{7}
  \begin{center}
  \tabcolsep=0.16cm	
        
   \begin{tabular}{cccc|ccccc|c}
      \multicolumn{2}{c}{Type} & $T_c$ & Chan.
                        & $N_{B=1}$
                        & $N_{B=2}$
                        & $N_{B=3}$
                        & $N_{B=4}$
                        & $N_{B \geq 5}$
                        & $B_{\operatorname{Max}}$ \\
			&  & ($\unit[]{ms}$) & ($X$)
			& \multicolumn{5}{c|}{} \\

      \hline \hline
      \multirow{12}{*}{\begin{sideways}Multicast\end{sideways}} 
			& \multirow{6}{*}{\begin{sideways}Weekend\end{sideways}} & 
              & $A$   &  5961 &  33 &  1 & 0 & 0 & 3 \\
       & & 10 & $B$   & 15288 & 526 & 47 & 3 & 1 & 5 \\
       & &    & $A\!B$  &    24 &   0 &  0 & 0 & 0 & 1 \\
      \cline{3-10}
       & &     & $A$  &  616 &   6 &  0  & 0 & 0 & 2 \\
       & & 100 & $B$  & 4387 & 256 & 15  & 1 & 0 & 4 \\
       & &     & $A\!B$ &   25 &   0 &  0  & 0 & 0 & 1 \\
      \cline{2-10}
        & \multirow{6}{*}{\begin{sideways}Weekdays\end{sideways}} & 

              & $A$   & 14441 &  640 & 178 & 50  & 21 & 9 \\
       & & 10 & $B$   & 73992 & 4523 & 725 & 121 & 48 & 6 \\
       & &    & $A\!B$  &   225 &    2 &   0 & 0   & 0  & 2 \\
  
     \cline{3-10}
       & &     & $A$  & 4047 & 139 & 19 & 3 & 1 & 5 \\
       & & 100 & $B$  & 8257 & 560 & 48 & 5 & 1 & 5 \\
       & &     & $A\!B$ &  148 &   0 &  0 & 0 & 0 & 1 \\
      \hline \hline

    \end{tabular}
  \end{center}
\end{table}

\begin{table*}[t]
  \caption{Experimental results concerning transmission latency and packet loss ratio\hspace{\textwidth}($\unit[2.4]{GHz}$ and $\unit[5]{GHz}$ bands, network manager disabled).}
  \label{tab:latency5GHz}
  \scriptsize
  \fontsize{6.0}{7}

  \begin{center}
  \tabcolsep=0.11cm	
    
    \begin{tabular}{cccc|cccc|cccccccc|c|cc}
      
      \multicolumn{2}{c}{Type} & $T_c$ & Chan. 
			& $\overline{d}$$^X$ & $\sigma_{d}^X$ & $d^X_{p99.99}$  & $d^X_{\operatorname{Max}}$ 
			& $\Upsilon_{d>1\mathrm{ms}}^X$ & $\hat{\Upsilon}_{d>1\mathrm{ms}}^{A\!C}$
			& $\Upsilon_{d>3\mathrm{ms}}^X$ & $\hat{\Upsilon}_{d>3\mathrm{ms}}^{A\!C}$
			& $\Upsilon_{d>10\mathrm{ms}}^X$ & $\hat{\Upsilon}_{d>10\mathrm{ms}}^{A\!C}$
			& $\Upsilon_{d>30\mathrm{ms}}^X$  & $\hat{\Upsilon}_{d>30\mathrm{ms}}^{A\!C}$
                        & $D_{KS}$
			& $\Upsilon_{L}^X$ & $\hat{\Upsilon}_{L}^{A\!C}$ \\
			&  & ($\unit[]{ms}$) & ($X$)
			& \multicolumn{4}{c|}{($\unit[]{ms}$)} 
			& \multicolumn{8}{c|}{(\textperthousand)}
                        & 
			& \multicolumn{2}{c}{(\textperthousand)} \\

      \hline \hline

      \multirow{6}{*}{\begin{sideways}Unicast\end{sideways}}  & \multirow{3}{*}{\begin{sideways}No Int.\end{sideways}} & 
              & $A$  & 1.21 & 1.46 & 35.14 & 79.08 & 250.0 & --     & 69.95 & --     & 3.511 & --     & 0.190 & -- &   & 0.000 & --     \\
       & & 10 & $C$  & 0.20 & 0.08 & 3.814 & 22.71 & 0.293 & --     & 0.152 & --     & 0.007 & --     & 0.000 & -- &   & 0.000 & --     \\
       & &    & $A\!C$ & 0.20 & 0.04 &  0.84 & 6.626 & 0.079 & 0.073 & 0.012 & 0.011 & 0.000 & 0.000 & 0.000 & 0.000 & 0.000 & 0.000 & 0.000 \\
      \cline{2-19}
        & \multirow{3}{*}{\begin{sideways}Int.\end{sideways}}  & 
              & $A$  & 1.32 & 1.71 & 41.55 & 135.5 & 284.9 & --     & 87.83 & --     & 4.839 & --     & 0.356 & -- &   & 0.000 & --     \\
       & & 10 & $C$  & 0.57 & 1.11 & 38.17 & 101.0 & 142.6 & --     & 17.60 & --     & 1.594 & --     & 0.190 & -- &   & 0.000 & --     \\
       & &    & $A\!C$ & 0.40 & 0.34 &  5.53 & 24.28 & 40.26 & 40.64 & 1.548 & 1.546 & 0.008 & 0.008 & 0.000 & 0.000 & 0.005 & 0.000 & 0.000 \\
      \hline \hline

      \multirow{6}{*}{\begin{sideways}Multicast\end{sideways}} & \multirow{3}{*}{\begin{sideways}No Int.\end{sideways}} & 
              & $A$  & 1.84 & 1.90 & 33.82 & 61.19 & 475.5 & --     & 156.4 & --     & 21.14 & --     & 12.53 & -- &   & 12.36   & --       \\
       & & 10 & $C$  & 0.89 & 0.08 &  3.99 & 24.98 & 31.24 & --     & 0.407 & --     & 0.230 & --     & 0.230 & -- &   & 0.230   & --       \\
       & &    & $A\!C$ & 0.88 & 0.07 &  2.26 & 21.19 & 14.39 & 14.86 & 0.066 & 0.064 & 0.006 & 0.005 & 0.003 & 0.003 & 0.047 & 0.00313 & 0.00284 \\

      \cline{2-19}
        & \multirow{3}{*}{\begin{sideways}Int.\end{sideways}} & 
              & $A$  & 2.28 & 2.95 & 46.13 & 545.9 & 607.1 & --     & 215.3 & --     & 31.65 & --     & 11.97 & -- &   & 11.50 & --     \\
       & & 10 & $C$  & 0.95 & 0.15 &  4.19 & 27.92 & 327.0 & --     & 66.02 & --     & 65.81 & --     & 65.81 & -- &   & 65.81 & --     \\
       & &    & $A\!C$ & 1.02 & 0.76 & 22.68 & 110.1 & 209.5 & 198.5 & 14.35 & 14.21 & 2.081 & 2.083 & 0.770 & 0.788 & 0.060 & 0.739 & 0.757 \\

      \hline \hline

    \end{tabular}
  \end{center}
\end{table*}

\subsection{Correlation between channels}
\label{sec:correlation_channels}
Estimates of some performance indices of a redundant link can be derived mathematically from statistical quantities measured separately on its channels.
Usually, any event (disturbance or interference) that causes a packet to be either dropped or delayed on one channel is not expected to improve the behavior of the other channel.
At best, it has no influence.
As a consequence, statistical independence among channels should be regarded as the theoretical ``best case''.
However, this is no longer true if some sources of interference are present that are not unrelated but act on channels at distinct instants (e.g., periodical network management activities whose occurrences on $A$ and $B$ are purposely displaced in time).
In these cases, the redundant link may perform better than expected.

\subsubsection{Packet Loss Ratio}
\label{sec:PLR}
Packets are lost in PoW only when they are dropped on both channels.
The probability $P_{l_i^{A\!B}}$ that packet $M_i$ is dropped on $\mathit{A\!B}$ is $P(l_i^A=1 \wedge l_i^B=1)$, and only in the case errors on $A$ and $B$ are statistically independent it equals $P_{l_i^A} \cdot P_{l_i^B}$.
As described in the previous sections, in the case the packet loss process is stationary and ergodic, measured values for the packet loss ratio $\Upsilon_L^X$ can be taken as reliable estimates of $P_{L^X}$ (and $P_{l_i^X}$).
This implies that statistical independence can be inferred by checking whether or not $\hat{\Upsilon}_L^{A\!B} \triangleq \Upsilon_L^A \cdot \Upsilon_L^B$ closely approximates $\Upsilon_L^{A\!B}$.

The computed value for $\hat{\Upsilon}_L^{A\!B}$ is reported in Table~\ref{tab:latency}, next to the experimental values for $\Upsilon_L^A$, $\Upsilon_L^B$, and $\Upsilon_L^{A\!B}$.
As can be seen by comparing $\hat{\Upsilon}_L^{A\!B}$ and $\Upsilon_L^{A\!B}$ in the lower part of the table, related to multicast transmissions (no unicast packets were in fact dropped), the assumption of statistical independence among channels almost satisfactorily holds.

\subsubsection{Deadline Miss Ratio}
\label{sec:DMR}
Under the same assumptions made above about the transmission process, latency on channel $X$ can be modeled as a random variable $D^X$.
Let $\overline{F}_{D^X}(h) \triangleq P(D^X>h)$ be its complementary cumulative distribution function (CCDF), i.e., the probability that the latency on channel $X$ is greater than $h$.
In this case, the probability that packet $M_i$ misses deadline $H$, because it is either lost or too late, coincides with the expression $P_{L^X}+(1-P_{L^X})\cdot \overline{F}_{D^{X}}(h)$ calculated for $h=H$, and can be estimated by the statistical quantity $\Upsilon_{d>H}^X$ measured in the experiments---see Eq.~(\ref{eq:dmr}).
When the fraction of lost packets is negligible, as on redundant links, $\Upsilon_{d>H}^{X}$ can be used to approximate the CCDF as well.

As happened for the packet loss ratio, independence between channels implies that $\Upsilon_{d>H}^{A\!B}$ could be satisfactorily approximated as $\hat{\Upsilon}_{d>H}^{A\!B} \triangleq \Upsilon_{d>H}^A \cdot \Upsilon_{d>H}^B$.
These values are reported in the third set of columns in Table~\ref{tab:latency}.
As can be seen, for multicast transmissions the theoretical estimates are quite close to the measured values (with some notable exceptions, though), whereas differences are larger in the unicast case.
Likely, this is due to the fact that channels are not truly independent (although they can be mostly considered as such for seamless channel redundancy, which is still effective).

\subsection{Adjacent channel interference}
\label{sec:ACI}
As a matter of fact, the experiments described
in the previous section revealed that some kind of phenomenon affected, in a non-negligible way, both channels of the redundant link at the same time.
Investigations permitted to identify the cause of such issues in the adjacent channel interference (ACI).
ACI is due to nearby transmitters, which ``bleed over'' to adjacent channels \cite{2008-IWCMC-ACM} even if their frequencies are spaced widely enough that, in theory, no interference should take place.
This may cause carrier sensing (CS) of a STA to believe that the BSS is not idle, hence uselessly delaying transmissions.
In our case, having the antennas of the two adapters close to each other (they are located about $\unit[5]{cm}$ apart) did cause interference on the CS mechanism, in spite of the fact the channels we selected were $1$ and $11$.

In order to try removing these dependencies, we performed a second set of experiments by operating the two adapters on different bands ($\unit[2.4]{}$ and $\unit[5]{GHz}$).
Increasing the distance between antennas was not an option: in fact, seamless link-level redundancy is mainly envisaged for the use in industrial devices (PLCs, sensors, and actuators), which typically fit in compact and small-sized enclosures.
Hence, we replaced the Wi-Fi equipment involved in measurements: although the testbed remained the same as in Fig.~\ref{fig:testbed}, dual-band adapters were installed in the PC while a dual-band AP was employed for channel $C$.

Since in our environment the traffic in the $\unit[5]{GHz}$ band was noticeably lower than in the ISM band, interfering load was purposely injected on $C$ by means of three additional STAs. 
Each interferer generates bursty traffic to mimic heavy and variable network activity.
The gap between bursts was chosen randomly according to an exponential distribution, and has mean duration equal to $\unit[1]{s}$. 
Each burst consists of $700$ unicast packets sent every $\unit[500]{\mu s}$ and with a payload size of $\unit[1500]{B}$. 
These interfering loads cause an additional traffic on channel $C$ that uses up (at least) $\unit[54]{\%}$ of the available bandwidth.

Results of the new trials are reported in Table~\ref{tab:latency5GHz}, with and without interfering traffic (labeled ``Int.'' and ``No Int.'', respectively).
As can be seen by checking the measured packet loss ratio and deadline miss ratio against the theoretical estimates, calculated as explained in Section~\ref{sec:correlation_channels}, a very good match is now achieved.
This confirms that, by taking suitable countermeasures in the design phase so as to prevent phenomena which jointly affect channels (due to either software, like the network manager, or hardware, like ACI), the benefits one may expect from redundant solutions in real operating conditions, in terms of the improvements they bring over single channels, are very close to what theory states.

\subsubsection{Distribution of latencies}
\label{Appendix}
In theory, the CCDF $\overline{F}_{D^{A\!B}}(h)$ for the latency $D^{A\!B}$, evaluated on packets correctly delivered on the redundant link, is related to the CCDFs $\overline{F}_{D^{A}}$ and $\overline{F}_{D^{B}}$ of its physical channels and their packet loss probabilities $P_L^A$ and $P_L^B$, through the equation:
\begin{align}
& \left(1-P_L^A \cdot P_L^B\right) \cdot \overline{F}_{D^{A\!B}}(h)=
   P_L^A \cdot P_R^B \cdot \overline{F}_{D^{B}}(h)+ \label{eq:teo_ccdf}\\
\nonumber
& +P_L^B \cdot P_R^A \cdot \overline{F}_{D^{A}}(h) +
   P_R^A \cdot P_R^B \cdot \overline{F}_{D^{A}}(h) \cdot \overline{F}_{D^{B}}(h)
\end{align}
where $P_R^X=1-P_L^X$ is the probability that packet transmission succeeds.
The term on the left side concerns the case when the packet is delivered on $A\!B$ but deadline $h$ is missed.
The first two terms on the right side refer to cases when packets are lost on one channel and arrive later than $h$ on the other, whereas the last term accounts for cases when both packets are received late.
In practice, because of the non-perfect independence, what can be found by using Eq.~(\ref{eq:teo_ccdf}) is just an estimate of $\overline{F}_{D^{A\!B}}(h)$, we denote $\hat{\overline{F}}_{D^{A\!B}}(h)$.

\begin{figure}[t]
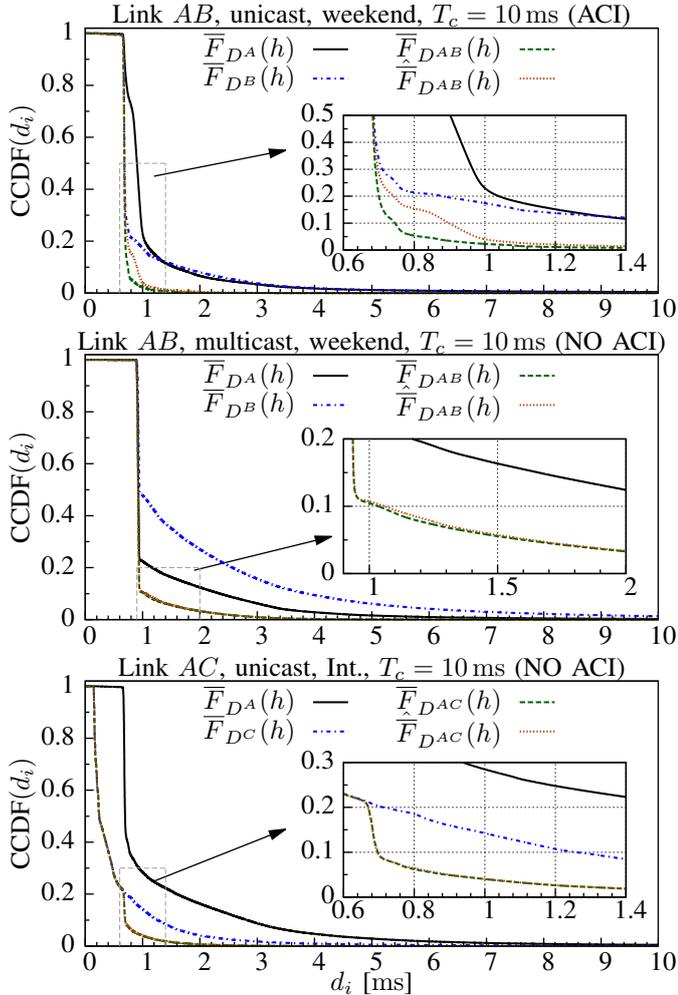

  \centering
  \include{FIG5-16-0457}
  \caption{Comparison between measured CCDFs of latencies and their estimates.}
  \label{fig:ccdf1}
\end{figure}

Starting from samples obtained in the experiments described in Section~\ref{sec:lab} (weekend condition and $T_c=\unit[10]{ms}$), we calculated the CCDFs for unicast and multicast traffic in the ISM band (channels $A$ and $B$).
They are reported in the two topmost plots in Fig.~\ref{fig:ccdf1}.
The plot at the bottom refers instead to unicast packets when the two channels are set on different bands ($A$ and $C$).
Besides the CCDFs of latencies measured on single channels, each plot shows the measured CCDF on the redundant link as well as its theoretical estimate obtained through Eq.~(\ref{eq:teo_ccdf}).

As can be seen, the measured and theoretical CCDFs on the redundant link $A\!B$ ($\overline{F}_{D^{A\!B}}(h)$ and $\hat{\overline{F}}_{D^{A\!B}}(h)$, respectively) for multicast packets are much closer than for the unicast case.
This is due to ACI affecting nearby adapters, which is absent in the multicast case as packets are sent by APs (located some meters apart).
ACI effects also disappear on the redundant link $AC$.
This is quite clear from the last plot, which shows a very good agreement between $\overline{F}_{D^{A\!C}}(h)$ and $\hat{\overline{F}}_{D^{A\!C}}(h)$.

\subsubsection{Reliability of CCDF estimation}
A quite interesting index to evaluate how much two CCDFs differ is the \textit{Kolmogorov-Smirnov distance} ($D_\mathit{KS}$), which represents the maximum absolute distance (error), computed on the y-axis, between them. 
By applying it to the measured CCDF for the redundant link and its theoretical estimate, we obtain:
\begin{equation}
D_\mathit{KS} = \sup_{h} \left| \overline{F}_{D^{A\!B}}(h) - \hat{\overline{F}}_{D^{A\!B}}(h) \right|
\end{equation}
A column has been added to Tables~\ref{tab:latency} and \ref{tab:latency5GHz} concerning this index.
For instance, for the plots in Fig.~\ref{fig:ccdf1} related to link $A\!B$, the maximum error for unicast packets ($D_\mathit{KS}=0.102$) is more than three times higher than for multicast packets ($D_\mathit{KS}=0.030$).
It is worth noting that the reported $D_\mathit{KS}$ values are quite pessimistic: in fact, the maximum error typically occurs near the point where CCDF plots start to fall, located just after the minimum transmission time (which corresponds to the best-case latency most packets experience in a well-designed system).
This means that, especially in the case of the link $AC$, independence between physical channels is actually better than this parameter suggests ($D_\mathit{KS}=0.005$).

$D_{KS}$ values have a practical use: for example, the probability $P(D^{A\!B}>\unit[1]{ms})$ that a received multicast packet has experienced a delay greater than $\unit[1]{ms}$ on the redundant link $A\!B$ can be obtained directly from the related theoretical CCDF, calculated from the latencies measured separately on the two channels, and corresponds to $\hat{\overline{F}}_{D^{A\!B}}(h) \mid_{h=\unit[1]{ms}} \pm D_\mathit{KS}$, i.e., $0.107 \pm 0.030$.

\section{Conclusion}
\label{Conclusion}
So far, Wi-Fi has been deemed not reliable enough for the use in time- and safety-critical systems, like those found in automated factory environments.
Seamless redundancy for IEEE 802.11 networks, which in the simplest cases can be obtained by layering PRP over conventional Wi-Fi equipment (PoW), permits to noticeably improve communication predictability.
Previous studies focused on performance evaluation of redundant wireless solutions (one of the most notable being Wi-Red), 
carried out using network simulation under the ground assumption that interference and disturbance on channels are statistically independent.

This paper describes a thorough experimental campaign, performed on a testbed made up of commercial Wi-Fi equipment, aimed at assessing to what extent such hypothesis is true.
Results point out that, when Wi-Fi networks are properly configured (i.e., DTIM, network manager, and ACI effects are prevented, see \cite{2016-ETFA-WiFi} for details), correlation between channels is so low that theoretical estimates for both the packet loss ratio and the deadline miss ratio on the redundant link provide close approximations of measured values, whose accuracy degree suits practical applications (in the experiments, they never differed by more than 10\%).

This shows, once again, that seamless redundancy is a viable option to improve Wi-Fi behavior.
Advantageously, there is no need to bring dramatic changes to the existing communication hardware, since the same radio and MAC blocks found in conventional wireless adapters can be exploited (at the limit, some slight modifications may be brought to the latter).
Moreover, also the interface to the upper layers remains unchanged.

It is important to remark that the results we found cannot be generically applied to every Wi-Fi network, without taking into account the specific traffic and environmental conditions.
Nevertheless, they are meaningful enough and justify the use of independent sources of disturbance and interference on redundant channels, at least for coarse analysis (both theoretical and simulation-based). 
As part of our future work, we plan to develop a software Wi-Red implementation and to assess its performance.

\section*{Acknowledgment}
The authors would like to thank Comau SpA, and in particular Dr.~Pietro Cultrona, Dr.~Michele Putero, and Dr.~Fulvio Rusin\`a, for their valuable support in carrying out measurements in the industrial plant.

\bibliographystyle{IEEEtran}
\bibliography{TII-16-0457}

\begin{IEEEbiography}%
[{\includegraphics[width=1in,height=1.25in,clip,keepaspectratio]
{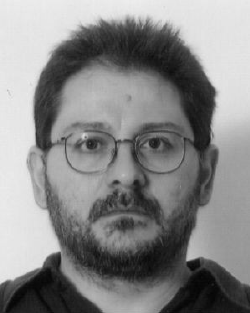}}]{Gianluca Cena} (SM'09) received the Laurea degree in electronic engineering and the Ph.D. degree in information and system engineering from the Politecnico di Torino, Italy, in 1991 and 1996, respectively.

In 1995 he became an Assistant Professor with the Department of Computer Engineering of the Politecnico di Torino. Since 2005 he has been a Director of Research with the National Research Council of Italy (CNR), and in particular with the Institute of Electronics, Computer and Telecommunication Engineering (IEIIT), Turin. His research interests include wired and wireless industrial communication systems, real-time protocols, and automotive networks. In these areas he has coauthored more than 120 technical papers and one international patent.

Dr. Cena served as Program Co-Chairman for the 2006 and 2008 editions of the IEEE International Workshop on Factory Communication Systems, and as Track Co-Chairman in six editions of the IEEE International Conference on Emerging Technologies and Factory Automation. Since 2009 he has been an Associate Editor of the \textsc{IEEE Transactions on Industrial Informatics}.
\end{IEEEbiography}

\begin{IEEEbiography}%
[{\includegraphics[width=1in,height=1.25in,clip,keepaspectratio]
{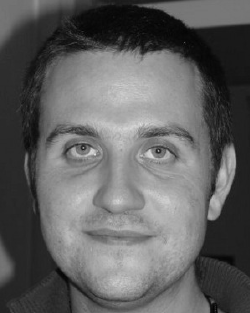}}]{Stefano Scanzio} (S'06-M'12) received the Laurea and Ph.D. degrees in Computer Science from Politecnico di Torino, Torino, Italy, in 2004 and 2008, respectively.

He was with the Department of Computer Engineering, Politecnico di Torino, from 2004 to 2009, where he was involved in research on speech recognition and, in particular, he has been active in classification methods and algorithms. Since 2009, he has been with the National Research Council of Italy (CNR), where he is a tenured Researcher with the Institute of Electronics, Computer and Telecommunication Engineering (IEIIT), Turin.

Dr. Scanzio teaches several courses on Computer Science at Politecnico di Torino. He has authored and co-authored several papers in international journals and conferences in the area of industrial communication systems, real-time networks, wireless networks and clock synchronization protocols.
\end{IEEEbiography}

\begin{IEEEbiography}%
[{\includegraphics[width=1in,height=1.25in,clip,keepaspectratio]
{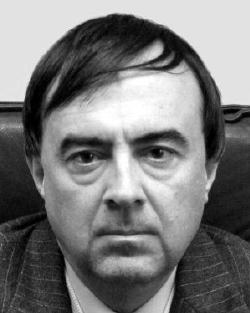}}]{Adriano Valenzano} (SM'09) received the Laurea degree in electronic engineering from Politecnico di Torino, Torino, Italy, in 1980. He is Director of Research with the National Research Council of Italy (CNR). He is currently with the Institute of Electronics, Computer and Telecommunication Engineering (IEIIT), Torino, Italy, where he is responsible for research concerning distributed computer systems, local area networks, and communication protocols. He has coauthored approximately 200 refereed journal and conference papers in the area of computer engineering.

Dr. Valenzano is the recipient of the 2013 IEEE IES and ABB Lifetime Contribution to Factory Automation Award. He also received, as a coauthor, the Best Paper Award presented at the Fifth and Eighth IEEE Workshops on Factory Communication Systems (WFCS 2004 and WFCS 2010). 

He has served as a technical referee for several international journals and conferences, also taking part in the program committees of international events of primary importance. Since 2007, he has been serving as an Associate Editor for the \textsc{IEEE Transactions on Industrial Informatics}.
\end{IEEEbiography}

\vfill

\end{document}